\documentclass{achemso}
\usepackage{xcolor}
\usepackage{amsmath,amssymb}
\usepackage{graphicx}

\usepackage{color}

\newcounter{lastnote}
\newenvironment{scilastnote}{%
\setcounter{lastnote}{\value{enumiv}}%
\addtocounter{lastnote}{+1}%
\begin{list}%
{\arabic{lastnote}.}
{\setlength{\leftmargin}{.22in}}
{\setlength{\labelsep}{.5em}}}
{\end{list}}

\title{Terahertz Excitonics in Carbon Nanotubes: \\Exciton Autoionization and Multiplication} 

\author{Filchito~Renee~G.~Bagsican}
\affiliation{Institute of Laser Engineering, Osaka University, Suita, Osaka 565-0871, Japan}
\altaffiliation{These authors contributed equally to this work}
\author{Michael~Wais}
\affiliation{Institute for Solid State Physics, TU Wien, 1040 Vienna, Austria}
\alsoaffiliation{Division of Physics and Applied Physics, School of Physical and Mathematical Sciences, Nanyang Technological University, Singapore}
\altaffiliation{These authors contributed equally to this work}
\author{Natsumi~Komatsu}
\affiliation{Department of Electrical and Computer Engineering, Rice University, Houston, Texas 77005, USA}
\author{Weilu~Gao}
\affiliation{Department of Electrical and Computer Engineering, Rice University, Houston, Texas 77005, USA}
\author{Lincoln~W.~Weber}
\affiliation{Department of Physics, Southern Illinois University Carbondale, Carbondale, Illinois 62901, USA}
\author{Kazunori~Serita}
\affiliation{Institute of Laser Engineering, Osaka University, Suita, Osaka 565-0871, Japan}
\author{Hironaru~Murakami}
\affiliation{Institute of Laser Engineering, Osaka University, Suita, Osaka 565-0871, Japan}
\author{Karsten~Held}
\affiliation{Institute for Solid State Physics, TU Wien, 1040 Vienna, Austria}
\author{Frank~A.~Hegmann}
\affiliation{Department of Physics, University of Alberta, Edmonton, Alberta T6G 2E1, Canada}
\author{Masayoshi Tonouchi}
\affiliation{Institute of Laser Engineering, Osaka University, Suita, Osaka 565-0871, Japan}
\author{Junichiro Kono}
\affiliation{Department of Electrical and Computer Engineering, Rice University, Houston, Texas 77005, USA}
\alsoaffiliation{Institute of Laser Engineering, Osaka University, Suita, Osaka 565-0871, Japan}
\alsoaffiliation{Division of Physics and Applied Physics, School of Physical and Mathematical Sciences, Nanyang Technological University, Singapore}
\alsoaffiliation{Department of Physics and Astronomy, Rice University, Houston, Texas 77005, USA}
\alsoaffiliation{Department of Material Science and NanoEngineering, Rice University, Houston, Texas 77005, USA}
\author{Iwao Kawayama}
\affiliation{Institute of Laser Engineering, Osaka University, Suita, Osaka 565-0871, Japan}
\alsoaffiliation{Graduate School of Energy Science, Kyoto University, Kyoto 606-8501, Japan}
\email{kawayama.iwao.3a@kyoto-u.ac.jp}
\author{Marco Battiato}
\affiliation{Division of Physics and Applied Physics, School of Physical and Mathematical Sciences, Nanyang Technological University, Singapore}
\email{marco.battiato@ntu.edu.sg}


\begin{document}
\maketitle
%



\date{}




\newpage


\begin{abstract}
Excitons play major roles in optical processes in modern semiconductors, such as single-wall carbon nanotubes (SWCNTs),\cite{MaultzschetAl05PRB2,WangetAl05Science,DukovicetAl05NL} transition metal dichalcogenides,\cite{ChernikovetAl14PRL,UgedaetAl14NM,HeetAl14PRL} and 2D perovskite quantum wells.\cite{IshiharaetAl89SSC,BlanconetAl18NC}  They possess extremely large binding energies ($>$100~meV), dominating absorption and emission spectra even at high temperatures.\cite{GaoetAl19ACS,Nishihara18}  The large binding energies imply that they are stable, that is, hard to ionize, rendering them seemingly unsuited for optoelectronic devices that require mobile charge carriers, especially terahertz emitters and solar cells.  
Here, we have conducted terahertz emission and photocurrent studies on films of aligned single-chirality semiconducting SWCNTs\cite{HeetAl16NN} and find that excitons autoionize, i.e., spontaneously dissociate into electrons and holes. This process naturally occurs ultrafast ($<$1~ps) while conserving energy and momentum.  The created carriers can then be accelerated to emit a burst of terahertz radiation when a dc bias is applied, with promising efficiency in comparison to standard GaAs-based emitters. Furthermore, at high bias, the accelerated carriers acquire high enough kinetic energy to create secondary excitons through impact exciton generation, again in a fully energy and momentum conserving fashion.  This exciton multiplication process leads to a nonlinear photocurrent increase as a function of bias.  Our theoretical simulations based on nonequilibrium Boltzmann transport equations, taking into account all possible scattering pathways and a realistic band structure, reproduce all our experimental data semi-quantitatively.  These results not only elucidate the momentum-dependent ultrafast dynamics of excitons and carriers in SWCNTs but also suggest promising routes toward terahertz excitonics despite the orders-of-magnitude mismatch between the exciton binding energies and the terahertz photon energies.
\end{abstract}

\newpage

The advent of modern low-dimensional semiconductor materials has opened up exciting possibilities for developing novel optoelectronic devices, utilizing their uniquely tunable optical properties which arise from unconventional degrees of freedom such as chirality, valley index, layer number, and twist angle.   In particular, single-wall carbon nanotubes (SWCNTs) provide ideal one-dimensional (1D) semiconductors whose properties depend on the chirality and diameter.\cite{WeismanKono19Book}  Recent seminal advances in enriching chiralities\cite{Zheng17TCC} and achieving alignment\cite{GaoKono19RSOS} on a macroscopic scale promise large-scale applications of SWCNT transistors, light-emitting diodes, and photodetectors.  Furthermore, the 1D nature of SWCNTs implies a highly restricted phase space for scattering, which in turn leads to ultrahigh carrier mobilities, suitable for high-frequency devices including terahertz (THz) emitters.\cite{HartmannetAl14Nano, Titova15}  Therefore, aligned and single-chirality SWCNT films are promising candidates for attaining electronic, THz, and optoelectronic functions on the same platform.

However, characteristic of low-dimensional semiconductors is their enormously enhanced exciton binding, compared to their bulk counterparts.  The typical exciton binding energies of semiconducting SWCNTs exceed 300~meV,\cite{MaultzschetAl05PRB2,WangetAl05Science,DukovicetAl05NL,WeismanKono19Book} as compared to 1-10~meV in typical bulk semiconductors.  This means that excitons in these semiconductors are very difficult to ionize, casting serious doubt about their abilities to generate THz radiation in response to an external electric field because excitons are charge-neutral particles.  Namely, to produce THz radiation efficiently, one needs to photogenerate charged free carriers that can be accelerated by an external field.  From this point of view, undoped low-dimensional semiconductors with huge exciton binding energies do not seem to be promising as optically excited THz emitters.

Here, we report on our finding that a photoconductive antenna (PCA) made from a film of aligned single-chirality semiconducting SWCNTs is an unexpectedly efficient THz emitter.  Without any optimization, we demonstrate that the produced THz intensities are only one order of magnitude lower than a state-of-the-art GaAs PCA. Furthermore, the CNT-based THz emitter can be easily fabricated on a variety of flexible substrates and is expected to be applied to compact THz analysis systems and wearable optoelectronics.
The THz intensity is resonantly enhanced through an interband exciton resonance, indicating the importance of excitons in the THz generation process.  To explain these observations, we developed a detailed microscopic model describing the strongly out-of-equilibrium dynamics and complex interplay of free carriers, excitons, photons, and phonons.  We took into account all energy and momentum conserving scattering processes with realistic band dispersions.  Through this, we shed light on a range of long-standing issues in SWCNT ultrafast dynamics.  Importantly, we show that $E_{22}$ excitons autoionize into free carriers right after the laser excitation.  We find multiexciton generation through exciton impact generation to be efficient at high bias.  This process explains the nonlinear bias dependence of the simultaneously measured DC photocurrent.

We fabricated a dipole-type PCA structure (Fig.\,1a) on top of a highly-aligned and chirality-enriched (6,5) SWCNT film deposited on a sapphire substrate (see Supplementary Material, SM). The nanotube alignment direction was aligned with the dipole gap during the fabrication process, so the applied DC electric field was parallel to the nanotube alignment direction (Fig.\,1a). The electrodes attached on the PCA structure were used to apply the DC bias and to measure the generated photocurrent at the same time (Fig.\,1a).  A separate low-temperature (LT)-GaAs PCA switch was used to detect the generated THz radiation in a transmission configuration (Fig.\,1a).  
We used an optical parametric oscillator as the wavelength-tunable excitation source or pump.

\begin{figure}
 \includegraphics[width=\textwidth]{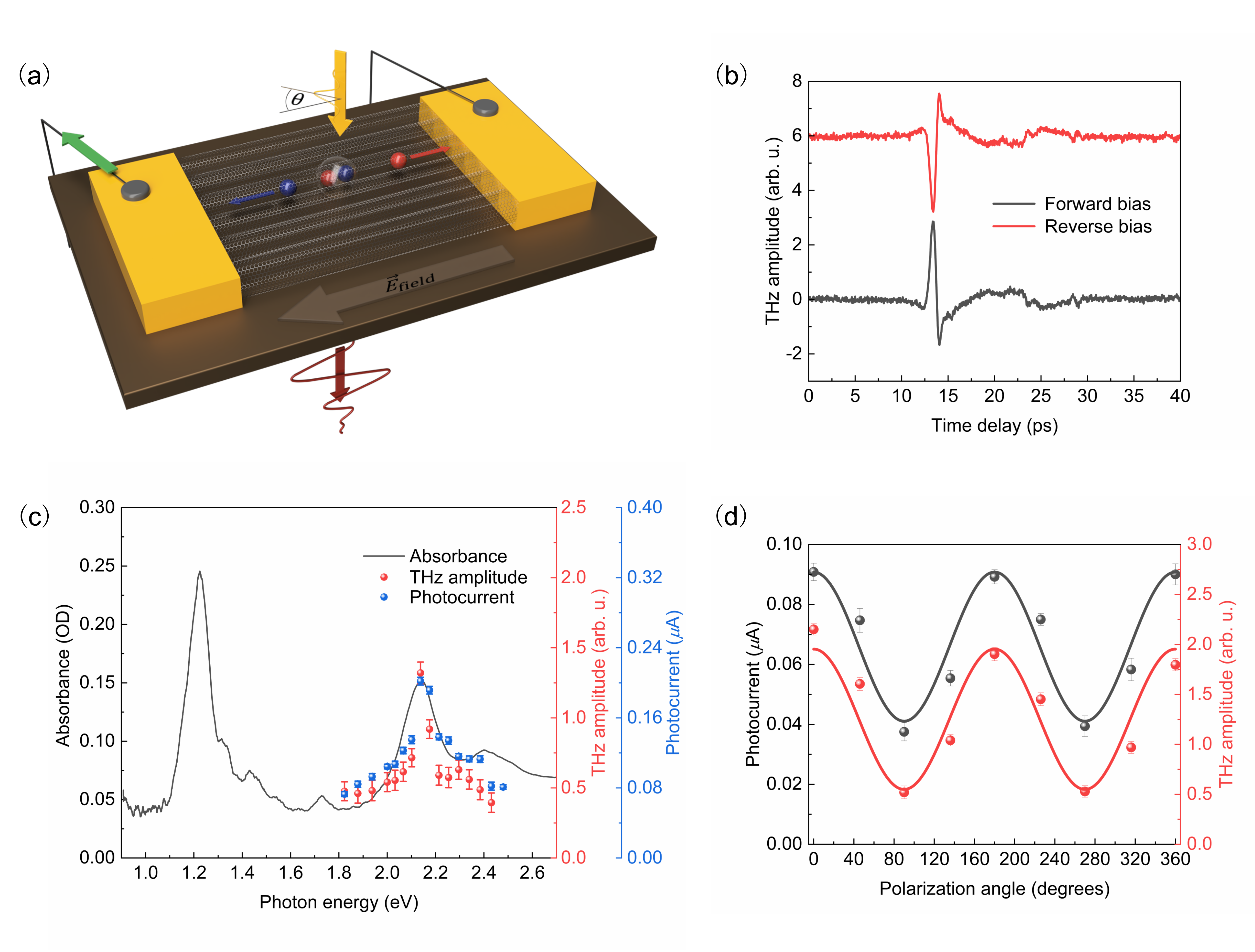}
 \caption{Schematic diagram of (a) CNT-based photoconductive antenna switch and experimental set-up. The CNTs are aligned with the direction of applied electric field. (b) THz emission waveforms at forward and reverse biases. (c) Absorbance spectrum for (6,5) CNT film plotted with the THz emission amplitude (red spheres) and photocurrent (blue spheres) as a function of photon energy. (d) THz amplitude and photocurrent as a function of angle between excitation femtosecond (fs) laser polarization and CNT alignment. 
}\label{fig:1}
\end{figure}

We excited the aligned SWCNTs around the $E_{22}$ exciton resonance for (6,5) SWCNTs.  Despite the fact that uncharged excitons are excited, a strong THz signal immediately appeared (Fig.\,1b).  The THz signal reversed its sign when the polarity of the applied biased was switched, confirming that the THz radiation is associated with a transient current parallel to the SWCNTs.  We then probed the dependence of both THz emission and photocurrent on the pump laser photon energy (Fig. 1c), as well as its polarisation (Fig. 1d). Both closely follow the behaviour of the $E_{22}$ exciton absorption, which peaks at $2.14\,\textrm{eV}$ and gets excited only by light polarized parallel to the nanotube axis. This further confirms that the THz generation process is initially triggered by the generation of excitons.

\begin{figure}
 \includegraphics[width=\textwidth]{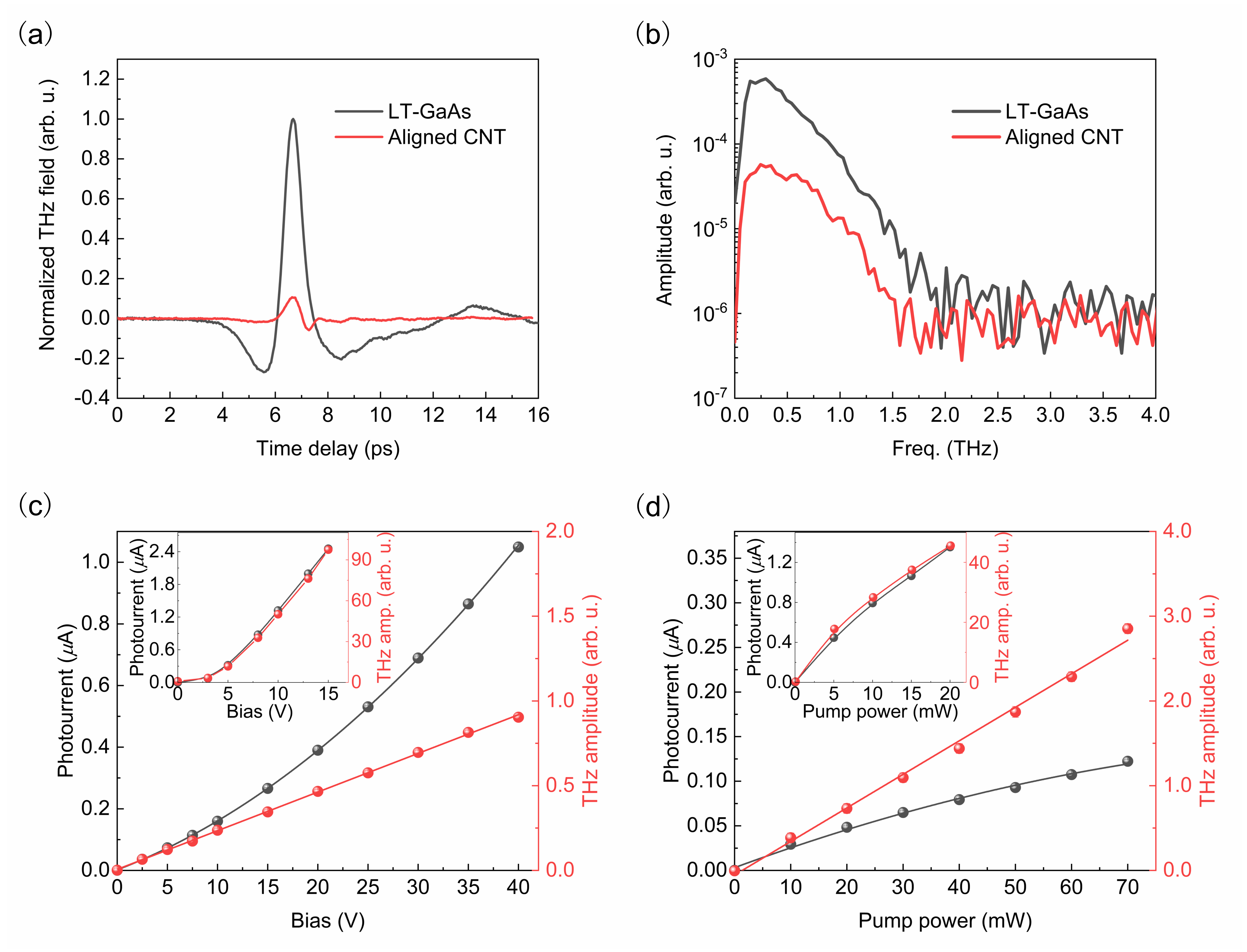}
 \caption{ Relative THz emission from CNT-based PCA and LT-GaAs-based PCA in (a) time and (b) frequency domains. Bias (c) and pump power (d) dependence of  photocurrent and THz emission from CNTs compared to same data from LT-GaAs-based PCA (insets).
}\label{fig:2}
\end{figure}

Figure 2a and 2b show the THz emission amplitude, as well as its spectrum, for our SWCNT device relative to that of a LT-GaAs-based PCA.  After considering the losses due to absorbance and reflectance (see SM), the emission efficiency (per thickness and absorbed power) of the SWCNT-based PCA is comparable within an order of magnitude to that of the LT-GaAs PCA. This is an excellent achievement for a prototype unoptimized device compared to the well-established THz emitter based on a conventional LT-GaAs PCA. 

We also carried out pump-intensity- and bias-dependent measurements. At a fixed pump intensity, both THz amplitude and photocurrent showed direct proportionality to the absorbance, with no apparent changes in the THz waveforms and spectra (see SM for details). However, we observed striking differences between THz emission and photocurrent in their pump power and bias dependences. As shown in Fig.\,2c and 2d, the THz amplitude shows a linear dependence on both the pump power and bias, whereas the photocurrent saturates at high pump powers and increases superlinearly with the bias.  This is in direct contrast with the LT-GaAs device wherein both THz emission and photocurrent always show the same behaviour. 

This difference in the bias dependence of THz and photocurrent suggests the emergence of new mechanisms for generating charged carriers in SWCNTs compared to typical PCAs. To provide a microscopic picture, we need to be able to accurately study the thermalization dynamics of several types of quasiparticles (electrons, excitons, photons, and phonons) within realistic band structure and in the presence of an external electric field. Describing the thermalization dynamics is particularly challenging: it requires full computation of far out-of-equilibrium scattering processes. The most critical complication is the precise accounting of energy and momentum selectivity of scattering events, which become exceptionally restrictive in 1D, and lead to critical deviation from typical Fermi liquid behaviours. Commonly used approximations\cite{Malic2013book,Reich2008,Knorr2009,Malic2013} either irreparably break the predictive power over the thermalization dynamics, particularly the ability to describe metastable states and many-timescales dynamics, or only allow for lower order processes (see SM for a complete description of numerical challenges).

\begin{figure}
 \includegraphics[width=\textwidth]{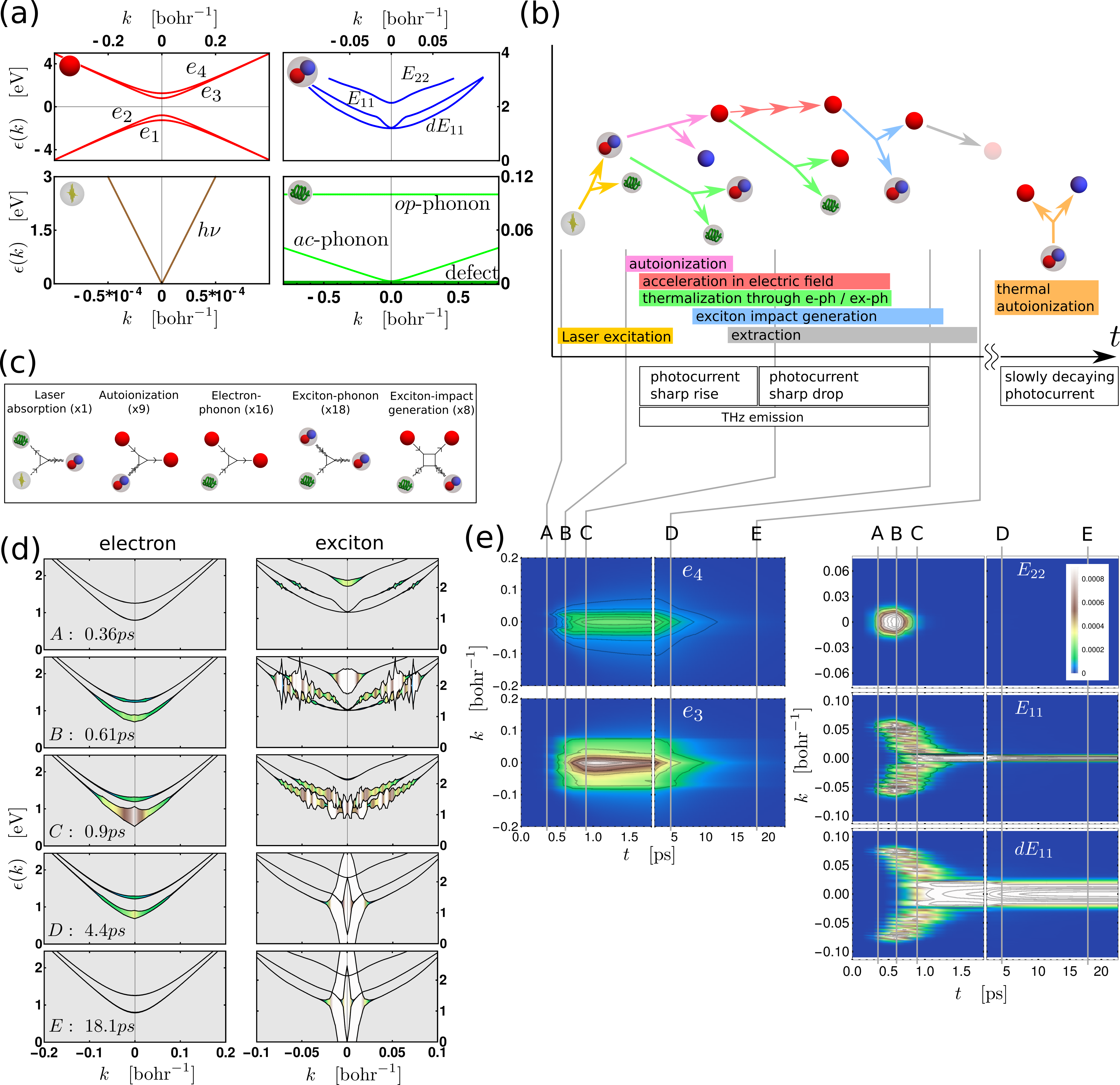}
 \caption{(a) Dispersions of the bands of all the included quasiparticles. (b) Qualitative description of the most influential scattering processes at different timescales. (c)  Schematic representation of all the types of the 52 scatterings included in the modelling. For more information see SM. (d) Time snapshots of the populations displayed over the dispersion: the thickness of the color bar as well as its colorscale represent the $k-$resolved population. The time propagation of the population in the remaining bands is calculated but not shown here. (e) Band resolved density plot of the quasiparticle population $f_n(k,t)$ for two of the electronic bands and all the excitonic bands. 
}\label{fig:3}
\end{figure}

Here we newly developed a numerical approach to the Boltzmann transport and scattering equation to simulate the microscopic processes leading to THz radiation and photocurrent generation in the aligned (6,5) SWCNT device.  We explicitly treated four electron bands\cite{CharlieretAl07RMP}($e_{1\dots4}$), three phonon\cite{LimetAl14NL}/defect bands ($ac$: acoustic phonon, $op$: optical phonon, $Imp$: impurity/defect), three exciton bands (one dark $E_{11}$, one bright $E_{11}$ and one bright $E_{22}$ exciton)\cite{DresselhausetAl07ARPC,MaultzschetAl05PRB2}, and a photon band ($h \nu$). Realistic dispersions were used for electrons, excitons, and photons, while effective phonon/defect bands were adopted to represent the large number of lattice modes and lattice imperfections (see Fig.\,3a, and SM for details). 
Quasiparticle populations are described as time ($t$)-, band ($n$)-, and momentum (${k}$)-dependent occupancies $f_{n} (k,t)$.  In thermal equilibrium, the populations are time-independent and given, in the case of fermions, by $f_{n}(k) = f_\mathrm{FD} (\varepsilon_{n}(k))$, where $f_\mathrm{FD}(\varepsilon)$ is the Fermi-Dirac distribution, and $\varepsilon_{n}(k)$ are the band dispersions. However, after the laser excitation, the quasiparticle populations can be extremely different from their equilibrium values, and the thermalization very often happens across a number of timescales, depending on the involved scattering channels. 

We took into account fifty two scattering channels (as well as their time reversal), including both three-leg scatterings (i.e., two particles in and one out) and four-leg scatterings (i.e., two particles in and two out, or three in and one out), as shown in Fig.\,3c.  All scatterings were calculated using the full, strongly out-of-equilibrium Boltzmann scattering integral (see SM). 
To our knowledge, this problem has so far been considered impossible to solve except in oversimplified systems or under strong approximations. Extensive methodological developments were necessary to overcome these limits (see SM).
Within the Boltzmann equation, we also took account of the fact that quasiparticles are subject to an external electric field due to the bias, which induces an acceleration proportional to the quasiparticle charge.  Due to the lack of spatial resolution at this stage of implementation of the numerical solver, the physical extraction of accelerated quasiparticles from the SWCNTs into the metallic leads was included via an effective population decay term (see SM).

The schematic diagram in Fig.\,3b describes the processes occurring at different timescales, and Figs.\,3d-3e show the populations of each band at selected time steps $A$-$E$. The laser excitation generates optically active $E_{22}$ excitons (Fig.\,3dA). They partially decay through scatterings with phonons\cite{ReichetAl05PRB,LauretetAl03PRL} into bright and dark $E_{11}$ excitons with similar energy but larger momentum (Fig.\,3dB). These quasiparticles are unaffected by the electric field and eventually decay to the bottom of the excitonic bands through scatterings with phonons (Fig.\,3dC).  A fraction of the excitons autoionize into electron-hole pairs in the conduction (valence) bands (Fig.\,3dB) through spontaneous dissociation\cite{Rumbles2015}. The free carriers are then accelerated by the electric field and shift in ${k}$ (Fig.\,3dC), inducing a rapid rise in current, which, in turn, generates THz radiation (Fig.\,4a).  The acceleration in momentum is limited by electron-phonon scatterings\cite{ManzonietAl05PRL,ParketAl04NL} (Fig.\,3dC), which prevent the electronic population from being accelerated indefinitely, simply leading to an asymmetry between positive and negative ${k}$ in the electronic population.  Eventually, the free carriers are physically extracted from the SWCNTs, thermalising in the metal contacts, leading to a drop in the electron and hole population and therefore a drop in the charge current (Fig.\,4a and Supplementary Material section 2.4). 

The simulated THz radiation (Figs.\,4a) and its spectrum (Figs.\,4b) are overall in good agreement with the experimental data. The time domain experimental THz signal shows a sharp negative peak directly after the main positive one originating from the switching off of the current. Our theoretical choice of the extraction mechanism leads to a broader and therefore smaller negative peak. At this stage we cannot make a claim on the details of the extraction mechanism since this discrepancy as well as the presence of further side peaks in the experimental THz radiation can be ascribed to suppression of low and high frequency components due to a variety of factors which were not incorporated in the simulations: inherent response of detector\cite{Tani_1997,Park1999,Duvillaret2001}, alignment and the effects of optical components along the THz propagation path from sample to detector\cite{Kuzel1999,Jepsen1996,Hattori_2002,VanRudd_2002}. All of these could easily distort the actual temporal and frequency profile of THz radiation from the sample. However, to verify the validity of our microscopic picture, we compare the effect of laser fluence (limited to the low fluence regime, see more in SM) and bias in the simulated dynamics to the measured ones. The population of the electronic conduction bands within 2~ps after excitation is proportional to the number of absorbed photons, which explains why the THz emission amplitude shows a linear dependence on the pump power.  An increase in the bias increases the induced asymmetry of the electronic distribution of both the conduction and valence bands. Since the number of autoionized free carriers is simply proportional to the laser power and not to the electric field, the induced current becomes proportional to the bias. Notice that we can exclude with certainty that the exciton break-up happens through field ionization, as the THz amplitude would, in that case, be superlinear with the bias.\cite{KumamotoPRL2014}

What remains still puzzling is the superlinear behaviour of the photocurrent with the applied voltage. One is usually tempted to assume the photocurrent and the THz emission to be simply different measurements of the same current, as our LT-GaAs control experiment seems to suggest (see inset in Fig.\,2c). However, there is a subtle difference between the two measurements: while the photocurrent is able to time integrate all the current generated in the PCA, only high-frequency components are emitted and measured as THz radiation.  Therefore, the two techniques provide insight into different timescales. By oversimplifying, one could say that the difference between the photocurrent and THz radiation is the amplitude of low-frequency currents. 

\begin{figure}
 \includegraphics[width=\textwidth]{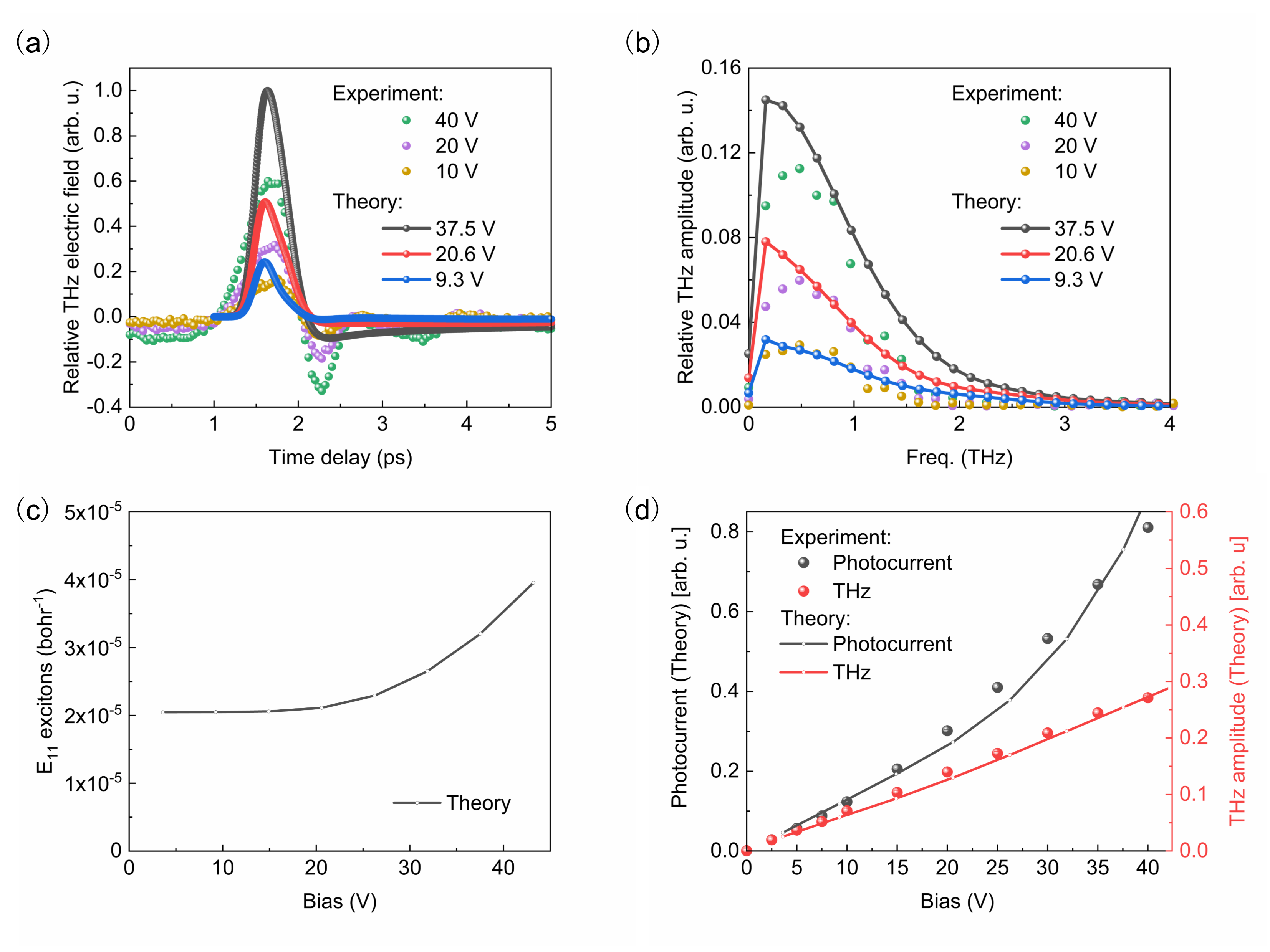}
 \caption{  Theoretical THz emission in (a) time and (b) frequency domain, compared to experimental measurements for three biases. 
 (c) Number of excitons left in the system at the last simulated time. (d) Comparison the computed peak amplitude of the THz emission and photocurrent with experimental results.
}\label{fig:4}
\end{figure}

The low energy excitons that are left in the system after the dynamics explained above (Fig.\,3dE) are not subject to the applied bias due to their zero net charge and therefore not accelerated and not subject to an important extraction from the SWCNTs. They survive at the bottom of the band and require longer times to annihilate through radiative or non-radiative processes (not included in the model). During this time, their high-energy tails will, due to the high temperature, slowly autoionize into free carriers (Fig.\,3b). These will generate a slow (i.e., low-frequency) electric current, which will not contribute to the THz emission, but will be measured at the electrodes as photocurrent. Such current will be proportional to the product of the number of excitons present in the system at the end of the initial thermalization (time E in Fig.\,3) and the bias. 

At a first sight, such current is expected to be simply linear in the bias. However, a careful analysis of all possible scatterings shows that an important four-leg scattering channel becomes active for electrons with an energy above a certain threshold: exciton impact generation (Fig.\,4c). An electron can scatter with an impurity (which provides momentum), lose energy, and generate an exciton in the process. Due to energy conservation, this scattering is allowed only for electrons with an energy larger than 1.2 eV above the band bottom. At low electric fields, only a small number of electrons can occupy such high energies. However, as the bias increases, this number grows. 
These electrons have therefore the chance of creating further excitons compared to the ones generated by the original optical excitation. 
As a consequence, the number of generated low-energy excitons has one component that is independent of the electric field, and another that instead grows with it (see Fig.\,4c). On a longer timescale (not included in our simulations), a fraction of these eventually autoionize into free carriers, which will again be accelerated by the electric field. Therefore, the photocurrent will be the sum of the picosecond current pulse and a low-frequency current generated by these autoionizing excitons. The low-frequency component of the photocurrent will be proportional to the product of the bias and the number of residual excitons at the end of the thermalization process (which depends partially on the bias itself). This explains the observed superlinear dependence of the amplitude of the photocurrent on the bias (see Fig.\,4d).


Notice that, since both the impulsive current responsible for the THz emission and the photocurrent are due to free carriers, measuring the same responses  after excitation of E11 excitons can provide a deep insight in the efficiency, the timescales as well as the mechanisms involved with the dissociation of these low energy excitons, compared to the dissociation channels of E22 excitons\cite{Beard2008,Soavi2015}.

%

Concluding, against expectations because of the large exciton binding energy, we produced, without any optimization, a prototype SWCNT-based THz emitter with an efficiency only an order of magnitude lower than highly optimized commercially available LT-GaAs -based emitters. With a suitable device design, it is possible to utilize the superior properties of SWCNT such as high thermal and chemical stability, to develop stable ultrabright sources. The strong confinement of current in a one-dimensional system could also offer better polarization control. In addition, their mechanical properties make them attractive for compact THz systems and wearable optoelectronics.

We further performed an unprecedentedly accurate theoretical analysis of the thermalization dynamics in these SWCNTs and explained the microscopic mechanisms behind the dissociation of excitons, free carriers dynamics in an applied electric field, and the conversion mechanisms between different quasiparticles. We identified the critical role of exciton impact generation in the ultrafast dynamics of SWCNTs causing a superlinear dependence of the photocurrent on the bias voltage.  Last but not least, our work shows that large-area films of aligned SWCNTs provide an exciting and promising playground for the study and use of the coexistence of positively and negatively charged as well as uncharged quasiparticles, as well as for the development of THz excitonics.

Supporting information is provided with this article. It contains, on the experimental side, information on device fabrication and experimental set-up, calculations for CNT degree of alignment and THz emission efficiency, and additional data at different pump energies. On the theoretical side, it covers a summary of the used band dispersion, a short introduction on the quantum Boltzmann equation, an analysis of each scattering channel used as well as the used scattering amplitudes, the description of carrier extraction and a brief overview of the numerical method. This material is available free of charge via the internet at http://pubs.acs.org.

\bibliographystyle{naturemag}
\bibliography{jun}

\begin{scilastnote}
\item [{\bf Acknowledgements:}] 
J.K.\ acknowledges support from the U.S.\ Department of Energy Basic Energy Sciences through Grant Number DEFG02-06ER46308 (optical spectroscopy), the U.S.\ National Science Foundation through Grant Number ECCS-1708315 (device fabrication), and the Robert A.\ Welch Foundation through Grant Number C-1509 (sample preparation). 
M.B. acknowledges Nanyang Technological University, NAP-SUG, for the funding of this research and the Austrian Science Fund (FWF) through Lise Meitner Grant No. M1925-N28.
M.W.~acknowledges the Austrian Science Fund (FWF) for funding through Doctoral School W1243 Solids4Fun (Building Solids for Function); and K.H. the FWF for support through project P30997. 
This work was partially supported by JSPS KAKENHI Grant Numbers JP18KK0140, JP18K18861, and JP19K15047, and by JSPS Core-to-Core Program, and Program for Promoting International Joint Research, Osaka University. I.K. acknowledges support from the Iketani Science and Technology Foundation.

\end{scilastnote}

\end{document}